\documentclass[sigconf, nonacm]{acmart}

\usepackage{listings}
\usepackage{tcolorbox}
\tcbuselibrary{skins,breakable}
\usepackage{tikz}
\usetikzlibrary{arrows.meta,positioning,shapes.geometric,calc}
\usepackage{booktabs}
\usepackage{enumitem}

\definecolor{code-bg}{HTML}{1E1E2E}
\definecolor{code-text}{HTML}{CDD6F4}
\definecolor{kw-blue}{HTML}{89B4FA}
\definecolor{kw-teal}{HTML}{94E2D5}
\definecolor{kw-green}{HTML}{A6E3A1}
\definecolor{str-peach}{HTML}{FAB387}
\definecolor{comment-gray}{HTML}{6C7086}
\definecolor{accent-blue}{HTML}{1E66F5}
\definecolor{accent-green}{HTML}{40A02B}
\definecolor{box-bg}{HTML}{EFF1F5}
\definecolor{box-border}{HTML}{7287FD}
\definecolor{proof-border}{HTML}{40A02B}
\definecolor{fact-border}{HTML}{179299}
\definecolor{result-border}{HTML}{40A02B}

\lstdefinelanguage{Cypher}{
  morekeywords={MATCH,WHERE,RETURN,WITH,CREATE,MERGE,SET,DELETE,DETACH,
    REMOVE,ORDER,BY,LIMIT,SKIP,UNWIND,FOREACH,CALL,YIELD,
    CASE,WHEN,THEN,ELSE,END,AND,OR,NOT,IN,AS,IS,NULL,
    TRUE,FALSE,LET,NEXT,FILTER,CYPHER,
    OPTIONAL,DISTINCT,UNION,ALL,
    ON,ERROR,BREAK,TRANSACTIONS,OF,ROW,
    REPEATABLE,ELEMENTS,NEXT},
  morekeywords=[2]{reduce,head,range,size,count,sum,collect,
    toInteger,split,trim,coalesce,toString,abs,allReduce},
  morekeywords=[3]{INC,JZDEC,HALT,JZDEC_ZERO,JZDEC_POS},
  sensitive=true,
  morestring=[b]',
  morestring=[b]",
  morecomment=[l]{//},
  morecomment=[s]{/*}{*/},
}

\lstdefinestyle{cypher}{
  language=Cypher,
  basicstyle=\ttfamily\fontsize{7pt}{8.5pt}\selectfont\color{code-text},
  keywordstyle=\color{kw-blue}\bfseries,
  keywordstyle=[2]\color{kw-teal},
  keywordstyle=[3]\color{kw-green},
  stringstyle=\color{str-peach},
  commentstyle=\color{comment-gray}\itshape,
  numbers=none,
  breaklines=true,
  tabsize=2,
  showstringspaces=false,
  frame=none,
  xleftmargin=0pt,
  aboveskip=0pt,
  belowskip=0pt,
}

\newtcolorbox{codeblock}{%
  colback=code-bg,colframe=code-bg,
  coltext=code-text,
  boxrule=0pt,
  borderline west={2pt}{0pt}{kw-teal},
  arc=2pt,
  left=5pt,right=5pt,top=4pt,bottom=4pt,
  breakable,
  before skip=0.6em,after skip=0.6em,
}

\newtcolorbox{claimbox}[1][Theorem]{%
  colback=box-bg,colframe=box-border,
  fonttitle=\bfseries\color{accent-blue},title=#1,
  boxrule=1pt,arc=2pt,left=5pt,right=5pt,top=4pt,bottom=4pt,
  before skip=0.6em,after skip=0.6em}

\newtcolorbox{proofbox}[1][Proof]{%
  colback={proof-border!5!white},colframe=proof-border,
  fonttitle=\bfseries\color{accent-green},title=#1,
  boxrule=1pt,arc=2pt,left=5pt,right=5pt,top=4pt,bottom=4pt,
  before skip=0.6em,after skip=0.6em,
  after upper={\hfill$\square$}}

\newtcolorbox{resultbox}[1][Result]{%
  colback={result-border!6!white},colframe=result-border,
  fonttitle=\bfseries\color{accent-green},title=#1,
  boxrule=1pt,arc=2pt,left=5pt,right=5pt,top=4pt,bottom=4pt,
  before skip=0.6em,after skip=0.6em}

\newtcolorbox{factbox}[1][Fact]{%
  colback=box-bg,colframe=fact-border,
  fonttitle=\bfseries\color{accent-blue},title=#1,
  boxrule=0.8pt,arc=2pt,left=5pt,right=5pt,top=4pt,bottom=4pt,
  before skip=0.6em,after skip=0.6em}

\title{Cypher is Turing-Complete: \\
  A Formal Proof via 2-Counter Machine Simulation}

\author{Pierre Halftermeyer}
\email{pierre.halftermeyer@neo4j.com}
\affiliation{%
  \institution{Neo4j}
  \country{France}
}

\begin{abstract}
We prove that Cypher~25, the graph query language of Neo4j, is
Turing-complete. The proof proceeds by showing that a single
\texttt{RETURN} statement using \texttt{reduce()}, \texttt{CASE}
expressions, and list comprehensions can simulate any 2-counter machine
(Minsky 1967). We address the bounded-step objection via two
complementary resolutions and present a third, graph-native simulation
using quantified path patterns (QPP) with \texttt{allReduce}, where the
machine \emph{is} the graph and computation \emph{is} path finding.
All three constructions are verified on a live Neo4j Aura instance.
\end{abstract}


\begin{document}

\maketitle

\section{The Claim}

\begin{claimbox}
Cypher~25, restricted to a single
\texttt{RETURN} with \texttt{reduce()}, list comprehensions, and
\texttt{CASE}, \textbf{is Turing-complete}.
\end{claimbox}

\textbf{Proof strategy.} By reduction:
$\text{TM} \xrightarrow{\text{Minsky}} \text{2CM} \xrightarrow{\text{this}} \text{Cypher}$.

\section{2-Counter Machines}

A \textbf{2-counter machine} $M = (Q, q_0, q_h, \delta)$ has a finite
state set~$Q$, initial state~$q_0$, halt state~$q_h$, and transition
function $\delta: Q \to \text{Instr}$ where each instruction is:

\begin{itemize}[leftmargin=1.2em,nosep]
  \item $\texttt{INC}(c, q')$ --- increment $c \in \{A,B\}$, goto $q'$
  \item $\texttt{JZDEC}(c, q_z, q_p)$ --- if $c{=}0$ goto $q_z$, else
        decrement and goto $q_p$
\end{itemize}

\begin{factbox}[Fact (Minsky 1967)]
For any TM~$T$, $\exists$ a 2-counter machine simulating~$T$.
\end{factbox}

\begin{center}
\begin{tikzpicture}[
  >=Stealth, scale=0.72, transform shape,
  state/.style={circle, draw=accent-green, fill=accent-green!15,
    thick, minimum size=0.8cm, font=\scriptsize\bfseries},
  halt/.style={circle, draw=accent-blue, fill=accent-blue!12,
    thick, minimum size=0.8cm, font=\scriptsize\bfseries},
  every edge/.style={draw, thick, ->, >=Stealth},
  elabel/.style={font=\tiny\ttfamily, inner sep=1pt},
]
  \node[state] (q0) at (0,0) {$q_0$};
  \node[state] (q1) at (3,0) {$q_1$};
  \node[state] (q2) at (0,-2.4) {$q_2$};
  \node[halt]  (q3) at (3,-2.4) {$q_3$};
  \draw[->] (q0) -- node[elabel, above] {INC(A)} (q1);
  \draw[->] (q1) -- node[elabel, right] {JZDEC(B), $B{>}0$} (q3);
  \draw[->] (q1) to[out=210,in=30] node[elabel, below, pos=0.45] {JZDEC(B), $B{=}0$} (q2);
  \draw[->] (q2) -- node[elabel, left] {INC(B)} (q0);
  \draw[->] (q3) edge[loop right] node[elabel, right] {HALT} (q3);
\end{tikzpicture}\\[2pt]
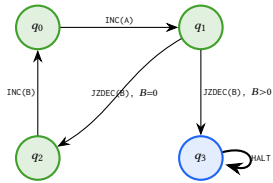
{\scriptsize Figure~1: Example 4-state program.}
\end{center}

\section{Encoding}

\textbf{Program.} Encode $\delta$ as a list of maps:

\begin{codeblock}
\begin{lstlisting}[style=cypher]
LET program = [
  {state:0, op:'INC',   counter:'A', next:1},
  {state:1, op:'JZDEC', counter:'B',
   q_zero:2, q_pos:3},
  {state:2, op:'INC',   counter:'B', next:0},
  {state:3, op:'HALT',  counter:'',  next:3}
]
\end{lstlisting}
\end{codeblock}

\textbf{State.} Configuration $(q, c_A, c_B)$ as map
\texttt{\{state:0, A:0, B:0\}}. Convention: \texttt{state=-1} means halted.

\section{The Simulation}

Each \texttt{reduce()} iteration executes one machine step.
The \texttt{head([v IN [e]| ...])} idiom provides let-binding
inside \texttt{reduce()} (where \texttt{LET} is unavailable).

\begin{codeblock}
\begin{lstlisting}[style=cypher]
CYPHER 25
LET program = [ /* ... */ ]
LET max_steps = 1000000

LET result = reduce(
  machine = {state:0, A:0, B:0},
  step IN range(1, max_steps) |
  CASE WHEN machine.state = -1
    THEN machine
  ELSE
   head([instr IN [program[machine.state]] |
    CASE instr.op
     WHEN 'INC' THEN
      CASE instr.counter
       WHEN 'A' THEN {state: instr.next,
         A: machine.A + 1, B: machine.B}
       WHEN 'B' THEN {state: instr.next,
         A: machine.A, B: machine.B + 1}
      END
     WHEN 'JZDEC' THEN
      CASE instr.counter
       WHEN 'A' THEN
        CASE WHEN machine.A = 0
         THEN {state: instr.q_zero,
               A: 0, B: machine.B}
         ELSE {state: instr.q_pos,
               A: machine.A-1, B: machine.B}
        END
       WHEN 'B' THEN
        CASE WHEN machine.B = 0
         THEN {state: instr.q_zero,
               A: machine.A, B: 0}
         ELSE {state: instr.q_pos,
               A: machine.A, B: machine.B-1}
        END
      END
     WHEN 'HALT' THEN
      {state: -1, A: machine.A, B: machine.B}
    END
   ])
  END
)
RETURN result
\end{lstlisting}
\end{codeblock}

\section{Correctness}

\begin{claimbox}[Claim]
After $k$ iterations of \texttt{reduce()}, \texttt{machine} equals
$M$'s configuration after $k$ steps.
\end{claimbox}

\begin{proofbox}[Proof by induction on $k$]
\textbf{Base.} $k{=}0$: \texttt{machine} $= \{0,0,0\} = M_0$.\enspace$\checkmark$

\textbf{Step.} Assume $\texttt{machine} = (q_k, a_k, b_k)$ matches $M$.
At $k{+}1$, \texttt{reduce()} looks up $\delta(q_k)$ and executes:

{\small
\begin{tabular}{@{}ll@{}}
$\texttt{INC}(A,q')$ & $\to (q', a_k{+}1, b_k)$ \enspace$\checkmark$ \\
$\texttt{JZDEC}(A,..)$, $a_k{=}0$ & $\to (q_z, 0, b_k)$ \enspace$\checkmark$ \\
$\texttt{JZDEC}(A,..)$, $a_k{>}0$ & $\to (q_p, a_k{-}1, b_k)$ \enspace$\checkmark$ \\
$\texttt{HALT}$ & $\to (-1, ..)$, identity after \enspace$\checkmark$ \\
\end{tabular}}

\noindent Symmetric for~$B$.
\end{proofbox}

\section{The Bounded-Step Objection}

The \texttt{reduce()} uses \texttt{range(1, max\_steps)} ---
a finite bound.

\subsection{Resolution via \texttt{IN TRANSACTIONS}}

While the pure functional version stays within a single expression,
the practical unbounded version uses graph storage for state persistence.

\begin{codeblock}
\begin{lstlisting}[style=cypher]
CYPHER 25
CREATE (:Machine {state:0, A:0, B:0});

UNWIND range(1, 9223372036854775807) AS step
CALL (step) {
  MATCH (m:Machine)
  WITH m,
    CASE WHEN m.state = -1 THEN 1/0
         ELSE $program[m.state]
    END AS instr
  SET m.state = CASE instr.op
    WHEN 'INC' THEN instr.next
    WHEN 'JZDEC' THEN CASE instr.counter
      WHEN 'A' THEN CASE WHEN m.A = 0
        THEN instr.q_zero ELSE instr.q_pos END
      WHEN 'B' THEN CASE WHEN m.B = 0
        THEN instr.q_zero ELSE instr.q_pos END
      END
    WHEN 'HALT' THEN -1 END,
  m.A = CASE
    WHEN instr.op='INC' AND instr.counter='A'
      THEN m.A+1
    WHEN instr.op='JZDEC' AND instr.counter='A'
      AND m.A > 0 THEN m.A-1
    ELSE m.A END,
  m.B = CASE
    WHEN instr.op='INC' AND instr.counter='B'
      THEN m.B+1
    WHEN instr.op='JZDEC' AND instr.counter='B'
      AND m.B > 0 THEN m.B-1
    ELSE m.B END
} IN TRANSACTIONS OF 1 ROW
  ON ERROR BREAK
\end{lstlisting}
\end{codeblock}

\textbf{Termination.} At halt, \texttt{state} becomes $-1$.
Next iteration: \texttt{CASE WHEN state=-1 THEN 1/0} fires
\emph{before} \texttt{\$program[-1]}, caught by \texttt{ON ERROR BREAK}.

\textbf{Note.} $2^{63} \approx 9.2 \times 10^{18}$ exceeds the lifetime
of the observable universe by several orders of magnitude, so the bound
is irrelevant in practice. True theoretical unboundedness would need an
external restart loop.

\subsection{Sufficiency Argument}

For TM~$T$ halting in $f(|w|)$ steps: set
$\texttt{max\_steps} = f(|w|)$. Cypher correctly decides every
decidable language --- \textbf{universal for total functions}.

\subsection{Resolution via Graph-Native QPP Traversal}

A third approach encodes the machine \emph{as} a graph and turns
computation into path finding. Each state becomes a node; each
instruction becomes a typed relationship (\texttt{INC},
\texttt{JZDEC\_ZERO}, \texttt{JZDEC\_POS}).
Only the program is persisted (as a small state-machine graph with
typed relationships); the counters are computed on-the-fly during
traversal and never written to the database.
A quantified path pattern (QPP) with \texttt{REPEATABLE ELEMENTS}
finds a path from \texttt{:Init} to \texttt{:Halt}, and
\texttt{allReduce} enforces counter constraints at every step,
pruning invalid branches eagerly.

\begin{codeblock}
\begin{lstlisting}[style=cypher]
CYPHER 25
CREATE (q0:State:Init {name: 'q0'})
CREATE (q1:State      {name: 'q1'})
CREATE (q2:State      {name: 'q2'})
CREATE (q3:State:Halt {name: 'q3'})
CREATE (q0)-[:INC        {c:'A'}]->(q1)
CREATE (q1)-[:JZDEC_ZERO {c:'B'}]->(q2)
CREATE (q1)-[:JZDEC_POS  {c:'B'}]->(q3)
CREATE (q2)-[:INC        {c:'B'}]->(q0)
\end{lstlisting}
\end{codeblock}

\vspace{-0.5em}

\begin{codeblock}
\begin{lstlisting}[style=cypher]
CYPHER 25
MATCH REPEATABLE ELEMENTS
  p = (init:Init)
    -[rels:INC|JZDEC_ZERO|JZDEC_POS]->
      {0, 9223372036854775807}
    (h:Halt)
WHERE allReduce(
  m = {A: 0, B: 0}, r IN rels |
  CASE
    WHEN r:INC AND r.c = 'A'
      THEN {A: m.A+1, B: m.B}
    WHEN r:INC AND r.c = 'B'
      THEN {A: m.A, B: m.B+1}
    WHEN r:JZDEC_POS AND r.c = 'A'
      THEN {A: m.A-1, B: m.B}
    WHEN r:JZDEC_POS AND r.c = 'B'
      THEN {A: m.A, B: m.B-1}
    ELSE m
  END,
  CASE
    WHEN r:JZDEC_ZERO AND r.c = 'A'
      THEN m.A = 0
    WHEN r:JZDEC_ZERO AND r.c = 'B'
      THEN m.B = 0
    WHEN r:JZDEC_POS AND r.c = 'A'
      THEN m.A >= 0
    WHEN r:JZDEC_POS AND r.c = 'B'
      THEN m.B >= 0
    ELSE true
  END
)
RETURN rels, length(p) AS steps
NEXT
LET final = reduce(m = {A:0, B:0},
    r IN rels |
    CASE
      WHEN r:INC AND r.c = 'A'
        THEN {A: m.A+1, B: m.B}
      WHEN r:INC AND r.c = 'B'
        THEN {A: m.A, B: m.B+1}
      WHEN r:JZDEC_POS AND r.c = 'A'
        THEN {A: m.A-1, B: m.B}
      WHEN r:JZDEC_POS AND r.c = 'B'
        THEN {A: m.A, B: m.B-1}
      ELSE m
    END
  )
RETURN steps, final.A AS ctrA, final.B AS ctrB
\end{lstlisting}
\end{codeblock}

\texttt{allReduce} folds over QPP relationships, updating an accumulator
and checking a predicate at each step. If the predicate returns
\texttt{false}, the path is pruned immediately. The predicate evaluates
on the \emph{post-update} accumulator: for \texttt{JZDEC\_POS} the
check is $\geq 0$ (not $> 0$), so decrementing from~$1$ to~$0$ is
accepted. Since \texttt{allReduce} does not expose the final
accumulator, \texttt{reduce()} in the \texttt{NEXT} stage re-derives
counter values.

\section{Required Primitives}

{\small
\begin{tabular}{@{}lp{4.2cm}@{}}
\toprule
\textbf{Primitive} & \textbf{Role} \\
\midrule
\texttt{reduce()} & Iteration (fold) \\
\texttt{CASE WHEN} & Branching \\
$+1$, $-1$, $=0$ & Counter ops \\
Map \texttt{\{s,A,B\}} & State repr. \\
\texttt{prog[i]} & $O(1)$ lookup \\
\texttt{head([..])} & Let-binding \\
\midrule
\multicolumn{2}{@{}l}{\emph{Additionally for QPP approach:}} \\
\texttt{REPEATABLE ELEMENTS} & QPP with revisitable nodes \\
\texttt{allReduce()} & Accumulator + pruning \\
\bottomrule
\end{tabular}}

\smallskip
\noindent\textbf{No graph operations. No APOC. No GDS.}

\section{Corollaries}

\begin{factbox}[Corollary 1: Undecidability of termination]
Since Cypher~25 can simulate any 2-counter machine, it is
\emph{undecidable} whether a given \texttt{reduce()} expression
will terminate.
(By Rice's theorem applied to the reduction from 2CM.)
\end{factbox}

\begin{factbox}[Corollary 2: Universality]
Cypher~25 can compute any computable function.
Any algorithm expressible as a Turing machine has an equivalent
\texttt{reduce()}-based Cypher query.
\end{factbox}

\begin{factbox}[Corollary 3: Minimality]
The proof uses \emph{no graph operations}: no \texttt{CREATE}, no
\texttt{MATCH}, no APOC, no GDS. Turing-completeness resides
in Cypher's expression language alone.
\end{factbox}

\section{Verified Execution}

All three approaches tested on Neo4j Aura (Neo4j 5.x, Cypher~25). Test program:
4 instructions, expected $\{-1, 2, 0\}$.

{\small
\begin{tabular}{@{}clccc@{}}
\toprule
\textbf{Step} & \textbf{Instr} & \textbf{St} & \textbf{A} & \textbf{B} \\
\midrule
0 & INC(A) & $q_0$ & $0{\to}1$ & 0 \\
1 & JZDEC(B), $B{=}0$ & $q_1$ & 1 & 0 \\
2 & INC(B) & $q_2$ & 1 & $0{\to}1$ \\
3 & INC(A) & $q_0$ & $1{\to}2$ & 1 \\
4 & JZDEC(B), $B{>}0$ & $q_1$ & 2 & $1{\to}0$ \\
5 & HALT & $q_3$ & 2 & 0 \\
\bottomrule
\end{tabular}}

\begin{resultbox}[\texttt{reduce()} result]
\texttt{\{A:2, B:0, state:-1\}} \enspace$\checkmark$
\end{resultbox}

\begin{resultbox}[\texttt{IN TRANSACTIONS} result]
\texttt{m.state=-1, m.A=2, m.B=0} \enspace$\checkmark$
\end{resultbox}

\begin{resultbox}[QPP \texttt{allReduce} traversal]
\texttt{steps=5, ctrA=2, ctrB=0} \enspace$\checkmark$
\end{resultbox}

\section{Complete Runnable Queries}

The following three queries are copy-paste ready for any instance
with Cypher~25 support. They encode the 4-state test
program from \S2 and can be adapted to any 2-counter machine by
editing the \texttt{program} list (or graph, for Approach~3).

\subsection{Approach 1: Pure \texttt{reduce()}}

A single \texttt{RETURN} --- no graph writes, no side effects.

\begin{codeblock}
\begin{lstlisting}[style=cypher]
CYPHER 25
LET program = [
  {state: 0, op: 'INC',   counter: 'A',
   next: 1},
  {state: 1, op: 'JZDEC', counter: 'B',
   q_zero: 2, q_pos: 3},
  {state: 2, op: 'INC',   counter: 'B',
   next: 0},
  {state: 3, op: 'HALT',  counter: '',
   next: 3}
]
LET max_steps = 1000000

LET result = reduce(
  machine = {state: 0, A: 0, B: 0},
  step IN range(1, max_steps) |
  CASE WHEN machine.state = -1
    THEN machine
  ELSE
    head([instr IN [program[machine.state]] |
      CASE instr.op
        WHEN 'INC' THEN
          CASE instr.counter
            WHEN 'A' THEN
              {state: instr.next,
               A: machine.A + 1, B: machine.B}
            WHEN 'B' THEN
              {state: instr.next,
               A: machine.A, B: machine.B + 1}
          END
        WHEN 'JZDEC' THEN
          CASE instr.counter
            WHEN 'A' THEN
              CASE WHEN machine.A = 0
                THEN {state: instr.q_zero,
                      A: 0, B: machine.B}
                ELSE {state: instr.q_pos,
                      A: machine.A - 1,
                      B: machine.B}
              END
            WHEN 'B' THEN
              CASE WHEN machine.B = 0
                THEN {state: instr.q_zero,
                      A: machine.A, B: 0}
                ELSE {state: instr.q_pos,
                      A: machine.A,
                      B: machine.B - 1}
              END
          END
        WHEN 'HALT' THEN
          {state: -1,
           A: machine.A, B: machine.B}
      END
    ])
  END
)
RETURN result
\end{lstlisting}
\end{codeblock}

\subsection{Approach 2: \texttt{IN TRANSACTIONS}}

Uses graph storage for state persistence.
The \texttt{1/0} guard fires \emph{before} any list indexing when
halted, caught by \texttt{ON ERROR BREAK}.

\begin{codeblock}
\begin{lstlisting}[style=cypher]
CYPHER 25
CREATE (:Machine {state: 0, A: 0, B: 0});
\end{lstlisting}
\end{codeblock}

\vspace{-0.5em}

\begin{codeblock}
\begin{lstlisting}[style=cypher]
CYPHER 25
LET program = [
  {state: 0, op: 'INC',   counter: 'A',
   next: 1},
  {state: 1, op: 'JZDEC', counter: 'B',
   q_zero: 2, q_pos: 3},
  {state: 2, op: 'INC',   counter: 'B',
   next: 0},
  {state: 3, op: 'HALT',  counter: '',
   next: 3}
]

UNWIND range(1, 9223372036854775807) AS step
CALL (step) {
  MATCH (m:Machine)
  WITH m,
    CASE WHEN m.state = -1 THEN 1/0
         ELSE program[m.state]
    END AS instr
  SET m.state = CASE instr.op
    WHEN 'INC' THEN instr.next
    WHEN 'JZDEC' THEN
      CASE instr.counter
        WHEN 'A' THEN
          CASE WHEN m.A = 0
            THEN instr.q_zero
            ELSE instr.q_pos END
        WHEN 'B' THEN
          CASE WHEN m.B = 0
            THEN instr.q_zero
            ELSE instr.q_pos END
      END
    WHEN 'HALT' THEN -1
    END,
  m.A = CASE
    WHEN instr.op = 'INC'
      AND instr.counter = 'A'
      THEN m.A + 1
    WHEN instr.op = 'JZDEC'
      AND instr.counter = 'A'
      AND m.A > 0 THEN m.A - 1
    ELSE m.A END,
  m.B = CASE
    WHEN instr.op = 'INC'
      AND instr.counter = 'B'
      THEN m.B + 1
    WHEN instr.op = 'JZDEC'
      AND instr.counter = 'B'
      AND m.B > 0 THEN m.B - 1
    ELSE m.B END
} IN TRANSACTIONS OF 1 ROW
  ON ERROR BREAK
\end{lstlisting}
\end{codeblock}

\begin{codeblock}
\begin{lstlisting}[style=cypher]
// Read result:
MATCH (m:Machine) RETURN m;
\end{lstlisting}
\end{codeblock}

\subsection{Approach 3: QPP + \texttt{allReduce}}

Graph setup and traversal query: see \S6.3.

\bibliographystyle{ACM-Reference-Format}

\end{document}